\documentclass[11pt]{iopart}
\usepackage{graphicx}
\usepackage{epsf,amsfonts,amssymb,amsbsy,wrapfig,cite}
\begin{document}
\title[Quantum spectrum of Kerr black hole: superstrings]
{A quantum mass-spectrum of Kerr black hole: superstrings}
\author{V.V.Kiselev*\dag}%\ddag}
%\fax{}
\address{*%\dag
\ Russian State Research Center ``Institute for
High Energy
Physics'', %\\
Pobeda 1, Protvino, Moscow Region, 142281, Russia}
\address{\dag\ %\ddag\
Moscow
Institute of Physics and Technology, Institutskii per. 9,
Dolgoprudnyi Moscow Region, 141700, Russia}
%\date{}
\ead{kiselev@th1.ihep.su}
\begin{abstract}
Using thermal quantization of geodesics confined under horizons
and reasonable conjecture on massless modes, we evaluate quantum
spectrum of Kerr black hole masses compatible with superstring
symmetries.
\end{abstract}
%\submitto{JCAP}
%\maketitle

%%%%%%%%%%%%%%%%%%%%%%%%%%%%%%%%%%%%%%%%%%%%%%%%%%%%%%%%%%%%%%%%%%

\section{Introduction}
In present paper we apply a thermal quantization approach
developed in \cite{Kiselev} to evaluation of quasi-classical
spectrum of masses in the case of rotating black hole described by
the Kerr metric. Main result is a linear dependence of angular
momentum on the black hole mass square, that reproduces a linear
trajectory characteristic for strings. The consideration is based
on thermal quantization of ratio of horizon areas. We extend
previous quantum set of ratios by including specific boundary
conditions \textit{a l\`{a}} Ramond and Neveu, Schwarz, that means
an involvement of supersymmetry.

Intercepts of trajectories are beyond the quasi-classical method,
and we need a consistent quantization of black holes including
notions of temperature, thermal ensemble and so on, which is still
not at hands. We offer a reasonable suggestion on a connection of
area ratio with spins of massless modes determining the
intercepts. The resulting mass-spectrum of quantum Kerr black hole
is compatible with trajectories of superstrings.

In section 2 we re-derive the Christodoulou-Ruffini mass formula
and use the thermal quantization to evaluate the spectrum of mass
for the Kerr black hole. Further we extend boundary conditions.
Section 3 is devoted to formulation of main conjecture on the
intercepts and massless modes. Then, we give corresponding quantum
spectra in agreement with symmetries of superstrings. In section 4
we describe dualities of quantized Kerr black holes. A connection
of thermal quantization with counting of microstates by Cardy
formula \cite{Cardy} is discussed in section 5. Final remarks are
given in section 6.

\section{Mass formula}
The Kerr--Newman metric of charged and rotating black hole can be
written in the form
\begin{equation}\label{KN}
    {\rm d}s^2_{\rm KN}=\frac{\Delta}{\Sigma}\,{\rm
    d}\omega_t^2-\frac{\Sigma}{\Delta}{\rm d}r^2-\Sigma\, {\rm
    d}\theta^2-\frac{\sin^2\theta}{\Sigma}\,{\rm d}\omega_\phi^2,
\end{equation}
with
\begin{equation}\label{defw}
    \begin{array}{rcl}
      {\rm d}\omega_t & = & {\rm d}t-a\,\sin^2\theta\,{\rm d}\phi,
      \\[2mm]
      {\rm d}\omega_\phi & = & (r^2+a^2)\,{\rm d}\phi-a\,{\rm d}t, \\
    \end{array}
    \qquad
    \begin{array}{rcl}
      \Sigma & = & r^2+a^2\,\cos^2\theta,
      \\[2mm]
      \Delta & = & (r-r_+)(r-r_-), \\
    \end{array}
\end{equation}
where the black hole parameters: mass $M$, charge $Q$, and angular
momentum $J$, are given by
\begin{equation}\label{MQJ}
    M=\frac{1}{2}(r_+ +r_-),\quad
    Q^2+a^2= r_+ r_-,\quad
    J =a\,M.
\end{equation}
Areas of two horizon surfaces at $r=r_\pm$ are equal to %given by
\begin{equation}\label{areas}
    {\cal A}_\pm =4\pi(r_\pm^2+a^2),
\end{equation}
and for convenience we introduce reduced areas
\begin{equation}\label{reduce}
    \mu_\pm^2\equiv\frac{{\cal A}_\pm}{4\pi}=r_\pm^2+a^2.
\end{equation}
Then relations (\ref{MQJ}) allow us express $r_-$ by
\begin{equation}\label{r-}
    r_-=\frac{Q^2+a^2}{r_+},
\end{equation}
and get the following equality
\begin{equation}\label{eq}
    Q^2+a^2+r_+^2=2 M r_+,
\end{equation}
that can be squared
\begin{equation}\label{sq}
    (Q^2+\mu_+^2)^2=4 M^2 (r_+^2+a^2-a^2)
\end{equation}
and rewritten in terms of $M$, $\mu_+$, and $J$ as
\begin{equation}\label{mass0}
    (Q^2+\mu_+^2)^2=4 M^2 \left(\mu_+^2-\frac{J^2}{M^2}\right).
\end{equation}
Thus, we arrive to the Christodoulou--Ruffini mass formula (see
discussion in short review \cite{Damour})
\begin{equation}\label{Mass}
    M^2=\frac{1}{4\mu_+^2}\left[(Q^2+\mu_+^2)^2+4J^2\right],
\end{equation}
which yields the mass of Kerr--Newman black hole in terms of its
horizon area, charge, and angular momentum. Moreover, one could
repeat the above derivation, excluding $r_+$ instead of $r_-$, so
that the final expression literally remains with no change, and we
deduce
\begin{equation}\label{MM}
    M^2=\frac{1}{4\mu_\pm^2}\left[(Q^2+\mu_\pm^2)^2+4J^2\right].
\end{equation}
Note, that (\ref{MM}) states a relation between $\mu_+$ and
$\mu_-$ at fixed charge and angular momentum.

\subsection{Thermal quantization}
Recently, we have considered radial geodesics completely confined
under horizons of black hole \cite{Kiselev}. We have found that
massive particles at such geodesics form a thermal statistical
ensemble with the Hawking temperature, and they are responsible
for the microscopic origin of black hole entropy. The geodesic
motion is characterized by a period in imaginary time $\beta$
inverse to the Hawking temperature. The periodicity leads to
`quasi-classical thermal quantization' or to thermodynamical
Bohr-orbits. A trajectory is ascribed to a \textit{winding number}
$n$, which gives a number of cycles per the full period. We have
introduced a phase variable $\varphi$ by rescaling the imaginary
time to the period of $2\pi$. In the case of two horizons, the
space-time of confined geodesics should be described by two
consistent maps. The consistency takes place, if the ratio of
horizon areas is quantized by
\begin{equation}\label{QuantArea}
    \frac{{\cal A}_+}{{\cal A}_-}=l,\qquad l\in \mathbb N.
\end{equation}
The winding numbers in two maps $n_\pm$ at ground level and the
\textit{loop parameter} $l$ are related as
\begin{equation}\label{nnl}
    n_+ =\frac{2 l}{l-1},\qquad n_-=\frac{2}{l-1},\qquad
    n_\pm\in\mathbb N,
\end{equation}
so that we have found the following admissible set of loops:
\begin{equation}\label{loops}
    l =\{1,2,3,\infty\}\quad \Rightarrow\quad
    n_+=\{\infty,4,3,2\}.
\end{equation}
The boundary values of $l=1$ and $l=\infty$ represent the extremal
and Schwarzschild black holes, respectively.

The thermal quantization yields
\begin{equation}\label{mu}
    \mu_+^2=l\,\mu_-^2,
\end{equation}
that allows us to express the black hole mass in terms of its
charge and angular momentum by exclusion of $\mu_\pm$ from
(\ref{MM}).

\subsection{Quasi-classical spectrum}
Let us investigate (\ref{MM}) at $Q=0$, i.e. the case of Kerr
black hole. Substituting (\ref{mu}) in (\ref{MM}) generically
results in
\begin{equation}\label{muJ}
    \mu_+^4 =4l\,J^2,
\end{equation}
so that excluding $\mu_+$, we deduce
\begin{equation}\label{M1}
    M^2=\frac{l+1}{2\sqrt{l}}\;J.
\end{equation}
Remarkably, at a leading Regge trajectory, a mass-spectrum of
quasi-classical (open/closed) string at $J\to \infty$ is
represented by the following linear dependence\footnote{For closed
string, we imply the ordinary substitution $\alpha^\prime\to
\alpha^\prime/2$.}:
\begin{equation}\label{open/closed}
    J = \alpha^\prime\,M^2,
\end{equation}
where $\alpha^\prime$ is a slope of trajectory or, equivalently,
the string tension. Therefore,
\begin{equation}\label{alp}
    \alpha^\prime_l=\frac{2\sqrt{l}}{l+1},
\end{equation}
which cover well know result for the extremal black hole $J=M^2$
at $l=1$.

Thus, we have found that the quasi-classical spectrum of Kerr
black holes repeats the mass-spectrum of string at the leading
trajectory with the following set of slopes:
\begin{equation}\label{slopes}
    \alpha^\prime =\{\alpha^\prime_l\}\textstyle
    =\left\{1,\,\frac{2}{3}\sqrt{2},\,\frac{1}{2}\sqrt{3},\,0\right\},
\end{equation}
in units $G=\hbar=c=1$, where $G$ is the Newton constant.

\subsection{Extension of boundary conditions}
As well know, a field $\Phi(\varphi)$ depending on a periodic
variable $\varphi\in [0,2\pi]$, can get two kinds of boundary
conditions: the bosonic and fermionic ones, i.e. the periodic and
anti-periodic boundary conditions. So,
\begin{equation}\label{FB}
    B:\; \Phi(0)=\Phi(2\pi),\qquad
    F:\; \Phi(0)=-\Phi(2\pi).
\end{equation}
Therefore, the total period can be twice greater.

Following Ramond, Neveu and Schwarz \cite{Ramond,NeveuSchwarz},
let us introduce two kinds of boundary conditions for two fields
normalized by
\begin{equation}\label{norm}
    \Phi_1(0)=\Phi_2(0),
\end{equation}
in the form
\begin{equation}\label{RNS}
\begin{array}{rl}
    \mbox{R:}& \Phi_1(2\pi)=+\Phi_2(2\pi),\\[3mm]
    \mbox{NS:}& \Phi_1(2\pi)=-\Phi_2(2\pi),
\end{array}
\end{equation}
so that two maps in the black hole interior are completely
consistent at periods $4\pi$ and $2\pi \tilde l$, respectively, so
that the ratio of areas is equal to
\begin{equation}\label{extend}
    \frac{{\cal A}_+}{{\cal A}_-}=\frac{\tilde l}{2}\ ,
    \qquad \tilde l\in \mathbb N,\;l>2.
\end{equation}
Requiring integer values for $n_\pm$, we find one additional loop
parameter: $l=3/2$. Hence, the complete set of loops is extended
to
\begin{equation}\label{complete}
    l=\left\{1,\frac{3}{2},2,3,\infty\right\}\quad\Rightarrow\quad
    \begin{array}{l}
    n_+=\{\infty,6,4,3,2\},\\[3mm]
    \alpha^\prime =\left\{1,\,\frac{2}{5}\sqrt{6},
    \,\frac{2}{3}\sqrt{2},\,\frac{1}{2}\sqrt{3},\,0\right\}.
    \end{array}
\end{equation}
Thus, the quasi-classical thermal quantization of Kerr black holes
leads to \textit{five} kinds of massive ground states and
\textit{five} values of slope for the linear trajectories at large
angular momenta of string. In this approximation, since the
maximal slope corresponds to the extremal black hole, one could
expect that the extremal black hole has a minimal mass among the
states at a given value of $J$.

The involvement of Ramond, Neveu and Schwarz boundary conditions
ordinary suggests a supersymmetry, so we suppose $J$ can have
half-integer values, too, that indicates properties of
superstrings.

\section{A reasonable conjecture: quantum spectrum}

We can reasonably suggest, first, that a complete quantization
leads to an ordinary renormalization, so that we put
\begin{equation}\label{J}
    J\quad\rightarrow\quad J-\alpha(0),
\end{equation}
that introduces a nonzero intercept of string trajectory
\begin{equation}\label{inter}
    J =\alpha^\prime\ M^2+ \alpha(0).
\end{equation}
The intercept represents spin of massless modes. So, let us
consider massless modes in the formalism of thermal quantization
of confined radial geodesics.

As we have already mentioned, the winding numbers listed in
(\ref{complete}) correspond to ground-state massive modes.
Geodesics of such modes have been described in \cite{Kiselev},
where we have pointed to that smaller values of winding numbers
should be ascribed to massless modes propagating on the horizon
surfaces. Therefore, excluding the case of extremal black hole, we
can list admissible winding numbers of massless modes at various
loops as follows:
$$ %%%\begin{equation*}
    \begin{array}{|c|c|c|}
    \hline
    \mbox{loop} & \mbox{massive} & \mbox{massless}\\
      l & n_+ & n \\
      \hline
      \vrule height4.7mm width0pt\frac{3}{2} & 6 & 5,4,3,2,1
      \\[0.7mm]
      2 & 4 & 3,2,1 \\
      3 & 3 & 2,1 \\
      \infty & 2 & 1 \\
      \hline
    \end{array}
$$ %%%%\end{equation*}
In quantization, the winding number is
dynamically conjugated to the periodic phase $\varphi$. A
quantum-mechanical description of such variables will be given
elsewhere. Formally, we can associate the winding number with the
number of spin states as
\begin{equation}\label{associate}
    n_+=2 s+1,
\end{equation}
where $s$ is the spin of massive particle. This analogy is
straightforward for the ground massive modes. Therefore,
identifying the spin of lightest massive mode $s$ with the total
angular momentum of black hole $J$, i.e. putting $J=s$, we can
determine the spin of lightest massive mode at the leading
trajectory of superstring. For instance, according to
(\ref{associate}) at $l=3/2$ the lightest massive state has spin
$s=5/2$, while massless modes have the following admissible set of
spins: $s=\{2,\ 3/2,\ 1,\ 1/2,\ 0\}$, which compose $N=2$
supermultiplet. Winding numbers of massless modes shown above
surprisingly  satisfy (\ref{associate}), however, this relation is
not applicable for \textit{irreducible} spin representations of
Poincare group for massless particles, and the analogy cannot be
strict\footnote{One could consider trajectories in infinitesimal
vicinity of horizon, that operates with massive modes. So, in such
the limit, massless modes will represent an appropriate set of
several irriducible components, which combine into irriducible
massive fields due to a Higgs-like mechanism.}. Nevertheless, the
construction allows us to associate spins of massless modes with
admissible set of string intercepts. So, the offered conjecture is
summarized in Table \ref{conject}.

\begin{table}[th]
  \centering
  \caption{The conjecture on the quantum spectrum.}\label{conject}

  \vspace*{2mm}
  \begin{tabular}{|c|c|c|c|c|}
    \hline
    loop & winding & massive & massless & slope \\
    & number & spin & spins & \\
    $l$ & $n_+$ & $s$ & $\alpha(0)$ & $\alpha^\prime$ \\
    \hline
    \vrule width0pt height5mm$\frac{3}{2}$ & $6$ & $\frac{5}{2}$ & $2,\ \frac{3}{2},\ 1,\ \frac{1}{2},\ 0$ &
    $\frac{2}{5}\sqrt{6}$ \\[1mm]
    \hline
    \vrule width0pt height5mm${2}$ & $4$ & $\frac{3}{2}$ & $1,\ \frac{1}{2},\ 0$ &
    $\frac{2}{3}\sqrt{2}$ \\[1mm]
    \hline
    \vrule width0pt height5mm${3}$ & $3$ & $1$ & $\frac{1}{2},\ 0$ &
    $\frac{1}{2}\sqrt{3}$ \\[1mm]
    \hline
    \vrule width0pt height5mm$\infty$ & $2$ & $\frac{1}{2}$ & $0$ &
    $0$ \\[1mm]
    \hline
  \end{tabular}
\end{table}

We see that the conjecture is not applicable to the extremal black
hole $l=1$ as well as to the Schwarzschild black hole $l=\infty$,
since the former has uncertain winding number of massive state,
and the latter should have zero angular momentum (or spin) and the
slope tending to zero. At $l=3/2$ the maximal supersymmetry is
$N=2$, while at $l=2,3$ it is $N=1$.

\begin{figure}[th]
  \setlength{\unitlength}{1.05mm}
  \begin{center}
  \begin{picture}(60,34)
  \put(0,0){\includegraphics[width=50\unitlength]{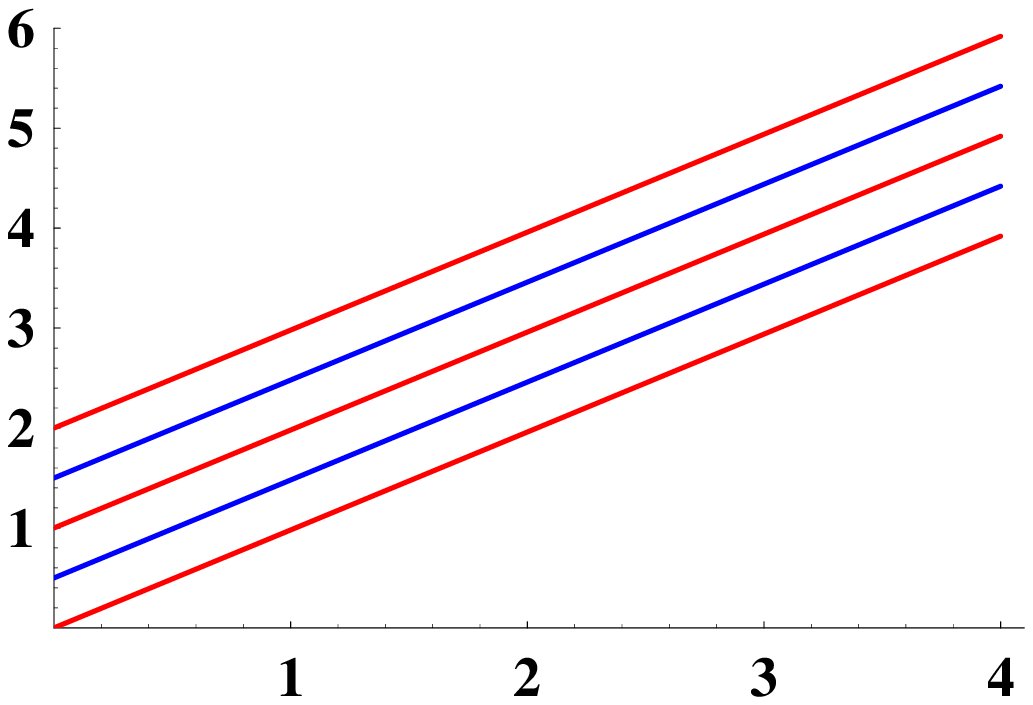}}
  \put(3,33){$J$}
  \put(50,2){$M^2$}
  \put(28,27){a)}
  \end{picture}
  \end{center}
  \begin{center}
  \begin{picture}(120,37)
  \put(0,0){\includegraphics[width=50\unitlength]{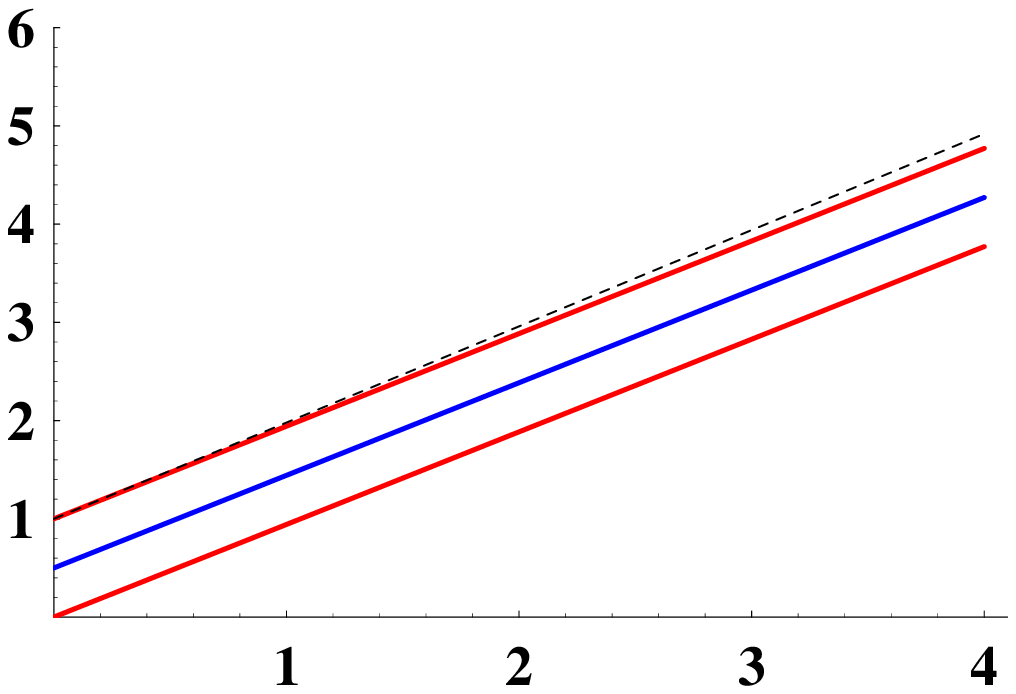}}
  \put(3,33){$J$}
  \put(50,2){$M^2$}
%  \end{picture}
%  \end{center}
%  \begin{center}
%  \begin{picture}(60,34)
  \put(60,0){\includegraphics[width=50\unitlength]{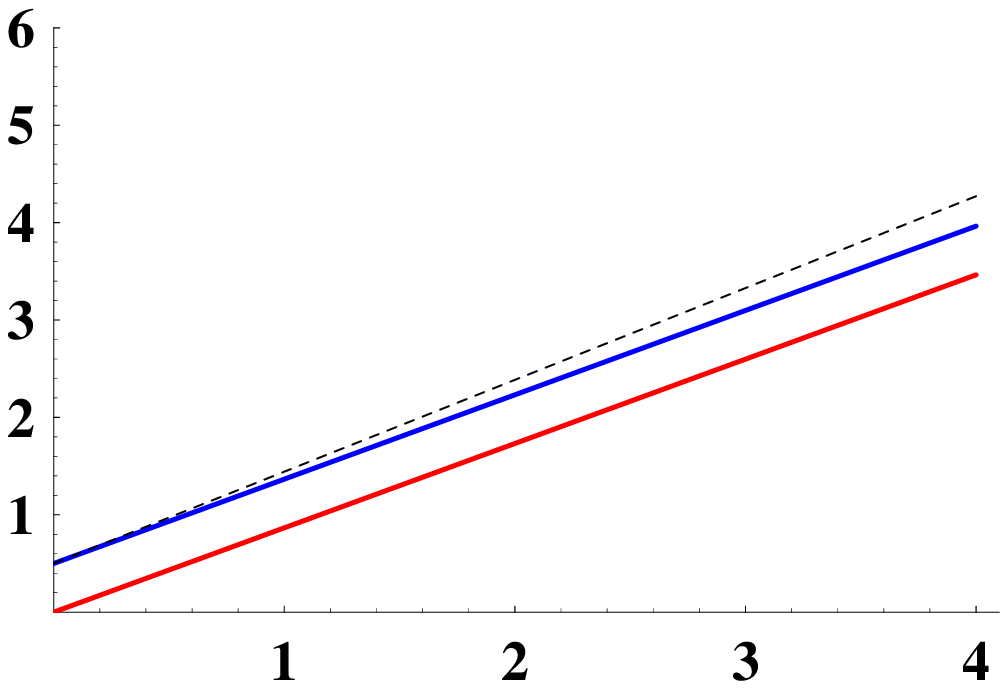}}
  \put(63,33){$J$}
  \put(110,2){$M^2$}
  \put(28,27){b)}
  \put(88,27){c)}
  \end{picture}
  \end{center}
  \caption{The trajectories of mass-spectrum at a) $l=3/2$, b) $l=2$,
  the dashed line represents the trajectory at $l=3/2$, c) $l=3$,
  the dashed line represents the trajectory at $l=2$.}\label{traject2}
\end{figure}

The trajectories of quantum mass-spectrum for the Kerr black hole
are shown in \mbox{Fig. \ref{traject2}}, where the mass squared is
given in units of $G^{-1}$. For comparison, we have also shown
trajectories with different slopes by dashed lines.

\begin{figure}[th]
  \setlength{\unitlength}{1.8mm}
  \begin{center}
  \begin{picture}(60,35)
  \put(0,0){\includegraphics[width=50\unitlength]{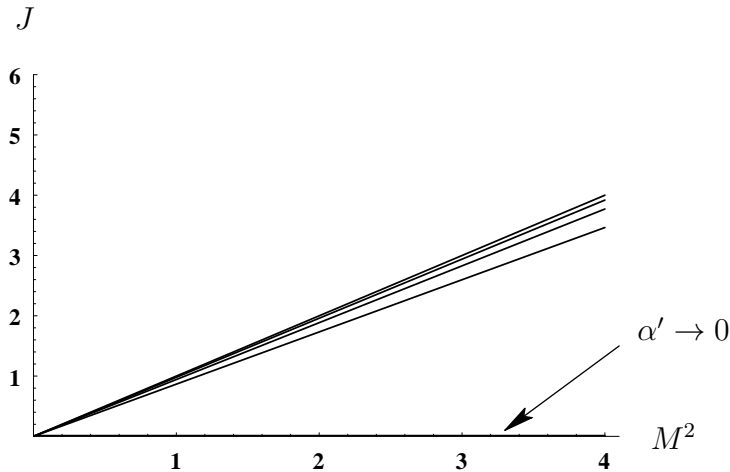}}
  \put(3,33){$J$}
  \put(50,2){$M^2$}
  \put(49,10){$\alpha^\prime\to 0$}
  \end{picture}
  \end{center}
  \caption{The trajectories of quasi-classical mass-spectrum with different slopes
  decreasing with the loop growth.}\label{quasi}
\end{figure}

Neglecting the renormalization, we return to the quasi-classical
spectrum of mass represented in Fig. \ref{quasi}, where the
extremal and Schwarzschild black holes are also shown.

Generically, we have five kinds of slopes. A question is whether
the number five accidentally coincides with five sorts of
superstrings or not.

\section{Dualities}

By construction, two horizons at $r_\pm$ enter the theory,
\textsl{symmetrically}. This property determines two kinds of
dualities.

First, the prescription of $r_+ > r_-$ is conventional. Therefore,
the mass spectra are not changed under the transformation
\begin{equation}\label{D1}
    l\;\leftrightarrow\; \frac{1}{l},
\end{equation}
which permutates $r_+\leftrightarrow r_-$. In this case, the
winding numbers at inner and outer horizons interchange:
$n_+\leftrightarrow -n_-$.

Second, implicitly, we have supposed that an observer lives in the
outer space-time at $r>r_+$. One could observe the same Kerr black
hole from a point in the inner space-time at $r<r_-$. We expect
that the spectrum of mass should be identical for both such
observers. This fact means the second duality in the form
\begin{equation}\label{D2}
    l\;\leftrightarrow\; \frac{1}{l},
\end{equation}
which conserves the slope $\alpha^\prime$. However, the winding
numbers could change.

Quasi-classically, both dualities are valid, and they have
identical properties in terms of mass spectrum.

\section{Entropy}

For outer observers, the Bekenstein--Hawking entropy
\cite{Bekenstein,Hawking,Hawking2}
\begin{equation}\label{BH}
    {\cal S}= \frac{1}{4}\,{\cal A}_+
\end{equation}
can be expressed in terms of angular momentum $J$ and loop
parameter $l$, so that
\begin{equation}\label{S}
    {\cal S} = 2\pi\sqrt{J^2\ l},\qquad \mbox{at large } J,
\end{equation}
that could be compared with the Cardy formula \cite{Cardy} in
conformal field theory for asymptotic number of microstates ${\cal
N}$,
\begin{equation}\label{N}
    \ln{\cal N} = 2\pi \sqrt{\frac{1}{6}\ {\cal C}\ L_0},
\end{equation}
with given values of central charge ${\cal C}$ of Virasoro algebra
for generators of conformal symmetry $L_n$ and eigenvalue of
$L_0$. We can easily guess that
\begin{equation}\label{guess}
    {\cal C}\ L_0 =6\, J^2\ l.
\end{equation}
Moreover, one could expect that eigenvalue $L_0$ should represent
something like a number of waves per period, i.e. it would be
reasonable to set $L_0= l$, since the loop parameter $l$
determines the number of small periods in large one, hence, the
charge is given by the global characteristics of metric, ${\cal
C}= 6\, J^2$.

A question is: What is an explicit form of hidden conformal
symmetry?

\section{Discussion and Conclusion}
We have combined the thermal quantization of Kerr black hole with
the conjecture on the spin of black hole in order to get the
quantum mass-spectrum described by trajectories \textit{a l\`{a}}
superstrings. The conjecture does not cover important cases of
extremal and Schwarzschild black holes. These cases were described
in superstring theory in \cite{StromingerVafa+}. Another approach
suggests asymptotic conformal symmetry on the horizon surface
\cite{Carlip,Solovev,Carlip2,Strominger}, which allows one to
count microstates by Cardy formula. Remarkably, we have pointed to
the connection of thermal quantization method with the Cardy
formula, that, however, requires further investigations.

The calculated quantum spectrum of Kerr black holes can be
compared with that of obtained by Gour and Medved \cite{Gour} and
investigated by Das, Mukhopadhyay and Ramadevi \cite{Das} in the
framework of method offered by Barvinsky, Kunstatter and Das
\cite{Barvinsky}. So, they found the following quantum spectrum of
area:
\begin{equation}\label{GM}
    {\cal A}_+ =8\pi\left(n+J+\frac{1}{2}\right),\qquad n \in
    \mathbb N,
\end{equation}
in contrast to our quasi-classical expression
\begin{equation}\label{Amy}
    {\cal A}_+ =8\pi\;\sqrt{l}\ J.
\end{equation}
We can, at first, improve (\ref{Amy}) by including the `quantum
renormalization' in the form of following substitution:
\begin{equation}\label{subJ}
    J \to J -\alpha(0)+n,
\end{equation}
where $\alpha_n(0)=\alpha(0)-n$ represents intercepts of principal
trajectory ($n=0$) and daughter ones ($n\neq 0$). Then (\ref{Amy})
and (\ref{subJ}) include more states than (\ref{GM}), since
$\alpha_n(0)$ can be positive in contrast to corresponding
negative values in (\ref{GM}). After that, the only difference is
reduced to the factor of $\sqrt{l}$. Therefore, the quantization
of \cite{Gour,Das,Barvinsky} corresponds to the case of $l=1$,
i.e. the extremal black hole, only. Note, that both approaches
under comparison are not applicable to the extremal case itself.
However, one could think that they are theoretically valid in the
limit of $l\to 1$, so that the study above is reasonable. Thus,
our approach allow us to find more rich quantum spectrum of Kerr
black holes, which corresponds to superstrings with five different
slopes. However, such the procedure of (\ref{subJ}) leads to
point-like massless modes of black holes with spin $J=\alpha(0)$
because at the same spin we get ${\cal A}_+=0$.

We have seen that the conjecture offered has produced new
questions and challenges to the superstring theory. Let us make
two remarks. First, anyway, the dimension $D=4$ consideration
means effective superstrings, therefore, one does not expects
actual problems with dimensional anomalies (for instance, the
conformal anomaly). Second, the importance of Kerr black hole
raises a problem on a superstring motion in the Kerr metric (not
in Minkowski flat one) at least in the classical limit of $J\to
\infty$ (massive strings produce a curvature of space-time).

In conclusion, the main problem remains a consistent quantization
of black holes beyond the quasi-classical thermal approach.

\vspace*{3mm}
 This work is partially supported by the grant of the
president of Russian Federation for scientific schools
NSc-1303.2003.2, and the Russian Foundation for Basic Research,
grants 04-02-17530, 05-02-16315.

%\newpage

\end{document}